\documentstyle[12pt,epsfig]{article}
\newcommand{\beq}{\begin{equation}}
\newcommand{\eeq}{\end{equation}}
\newcommand{\beqn}{\begin{eqnarray}}
\newcommand{\eeqn}{\end{eqnarray}}
\newcommand{\stackM}{\stackrel{\scriptstyle >}{{ }_{\sim}}}
\newcommand{\stackm}{\stackrel{\scriptstyle <}{{ }_{\sim}}} 
\begin{document}
\thispagestyle{empty}
\def\pubnum{422}
\def\data{July, 1997}

\def\UUAABB{\hbox{
    \vrule height0pt width2.5in
    \vbox{\hbox{\rm 
     UAB-FT-\pubnum
    }\break\hbox{\data\hfill}
    \break\hbox{hep-ph/9707535\hfill} 
    \hrule height2.7cm width0pt}
   }}   
\hfill\UUAABB
\vspace{3cm}
\begin{center}
\begin{large}
\begin{bf}
IMPLICATIONS ON THE SUPERSYMMETRIC HIGGS SECTOR  
FROM TOP QUARK DECAYS AT THE TEVATRON 
\\
\end{bf}
\end{large}
\vspace{1cm}
Jaume GUASCH, Joan SOL{\`A}\\

\vspace{0.25cm} 
Grup de F{\'\i}sica Te{\`o}rica\\ 
and\\ 
Institut de F{\'\i}sica d'Altes Energies\\ 
\vspace{0.25cm} 
Universitat Aut{\`o}noma de Barcelona\\
08193 Bellaterra (Barcelona), Catalonia, Spain\\
\end{center}
\vspace{0.3cm}
\hyphenation{super-symme-tric sig-ni-fi-cant-ly ge-ne-ral
asso-cia-ted re-mar-ka-bly}
\hyphenation{com-pe-ti-ti-ve}
\hyphenation{mo-dels}
\begin{center}
{\bf ABSTRACT}
\end{center}
\begin{quotation}
\noindent
\hyphenation{ob-ser-va-bles com-pa-ti-bi-li-ty}
\noindent
From the $\tau$-lepton analysis of the charged Higgs decay of the top quark
at the Tevatron, 
$t\rightarrow H^+\,b\rightarrow \tau^+\,\nu_{\tau}\,b$,
it is possible to set rather stringent bounds on the 
$(\tan\beta,M_{H^\pm})$-plane, if one assumes that $H^\pm$ is a
charged member of a generic two-Higgs-doublet model. 
However, if we consider the possibility that $H^\pm$
is supersymmetric, then we find that the allowed 
region in the $(\tan\beta,M_{H^\pm})$-plane can be significantly modified
by the MSSM quantum corrections. 
Throughout our analysis we correlate
the top quark results with the limits imposed by 
radiative and semileptonic $B$-meson decays.
Remarkably, one can envision situations in the MSSM parameter
space where $H^\pm$ completely eludes those bounds,
i.e. a charged Higgs with a mass below the top quark mass could coexist
with essentially any value of $\tan\beta$. 
\end{quotation}
 
\baselineskip=6.5mm  
 
\newpage

The Minimal Supersymmetric Standard Model (MSSM)\,\cite{MSSM} is a
most tantalizing candidate to extend the quantum field
theoretical structure of the Standard Model (SM)
of the strong and electroweak interactions while keeping all the
necessary ingredients insuring internal consistency, such as
gauge invariance and renormalizability.
Furthermore, supersymmetric theories are 
the only tenable theoretical framework for aiming at an unified theory
of all interactions including gravity. In spite of the fact that
the sparticles themselves have so far escaped detection, SUSY
has always eschewed experimental refutation; well on the contrary, it
persists since its inception (roughly a quarter of century ago!) as
a challenging proposal for physics beyond the SM.
For one thing the MSSM remains consistent with all known high precision
data at a level comparable to the SM\,\cite{WdeBoer,CPOK}, a feature which is
unparalleled by any alternative paradigm of the basic
particle interactions of nature.

In this note we address the possibility of seeing indirect effects of
SUSY through the interplay between top quark and Higgs boson physics.
A particularly distinctive part of the  field content of the MSSM is the Higgs 
sector, which is extended to contain two doublet scalar
fields leading to five physical states, namely two charged Higgs 
pseudoscalar bosons, $H^{\pm}$,
one neutral CP-odd boson, $A^0$, and two neutral
CP-even Higgs states, $h^0$ and $H^0$ ($M_{h^0}< M_{H^0}$)\cite{Hunter}.
Probing this extended Higgs sector can be very useful to search for SUSY,
due to the special constraints imposed by the underlying symmetry.
In this respect, much work has been done mainly for the
neutral Higgs bosons\,\cite{Hunter}, and
especially in connection with the lightest CP-even state, $h^0$, which could 
mimic the SM Higgs boson in the allowed mass range
$M_{h^0}\stackm 130\,GeV$ -- i.e. the one obtained after including the MSSM
quantum corrections\,\cite{Higgsloop}. Thus, in principle, finding a
heavy ($>130\,GeV$) neutral Higgs scalar with similar couplings as
the SM Higgs boson would be a natural disprove of the MSSM. Nonetheless,
even if such a state were the first one to be found, it would be far from
clear whether it fully behaved as a SM Higgs boson, at least not before a lot
of   
precision data  were collected on the new particle. On the other hand, if
a charged Higgs boson, $H^{\pm}$, happened to exist, its identification 
should in principle be easier (if it were light enough)
and it would constitute an undeniable sign of physics beyond the SM.
However, also in this case its nature would not be immediately apparent.
For, in the absence of
other Higgs bosons (for example, the neutral ones of the MSSM), we
could not clearly distinguish whether we would be dealing with a SUSY charged
Higgs or e.g. with some charged member of a generic
two-Higgs-doublet model ($2HDM$). 

Along this vein the CDF Collaboration at the Tevatron
has carried out direct searches for charged Higgs production 
in $p\,\bar{p}$ collisions at
$\sqrt{s}=1.8\,TeV$\,\cite{Conway,CDF}. 
In these studies one is concerned with the final configurations
$t\bar{t}\rightarrow H^+\,H^-\,b\,\bar{b}, 
W^+\,H^-\,b\,\bar{b},H^+\,W^-\,b\,\bar{b}$.
The latter would differ from that of the standard model,
$t\bar{t}\rightarrow W^+\,W^-\,b\,\bar{b}$, by
an excess of states with one (or two) $\tau$-lepton ``jets'' 
(i.e. usually tagged in the hadronic decay mode) and two b-quarks 
and large missing transverse energy associated to
the decays $H^+\rightarrow \tau^+\,\nu_{\tau}$ and/or 
$H^-\rightarrow \tau^-\,\bar\nu_{\tau}$.
To grasp a hint of the relative importance of these events, notice
the following rates 
(where for illustration purposes we have just
kept the relevant Yukawa couplings): 
\beq
{\Gamma^{(0)}(t\rightarrow H^+\,b)\over \Gamma^{(0)}(t\rightarrow W^+\,b)}=
{\left(1-{M_{H}^2\over m_t^2}\right)^2\,
\left[{m_b^2\over m_t^2}\,\tan^2\beta+\cot^2\beta\right]
\over
\left(1-{M_W^2\over m_t^2}\right)^2\,\left(1+2{M_W^2\over m_t^2}\right)}\,,
\label{eq:ratioHW} 
\eeq     
and (at large $\tan\beta$)
\beqn  
{\Gamma^{(0)}(H^{+}\rightarrow\tau^{+}\nu_{\tau})\over
\Gamma^{(0)}(H^{+}\rightarrow c\bar{s})}&=&\frac{1}{3}
\left(\frac{m_{\tau}}{m_c}\right)^2
{\tan^2\beta\over
(m^2_s/ m^2_c)\tan^2\beta+\cot^2\beta}
\rightarrow \frac{1}{3}\left(\frac{m_{\tau}}{m_s}\right)^2
>10\,.
\label{ratiotaucs}
\eeqn
We see from eq.(\ref{eq:ratioHW}) that if $M_H$ (the mass of
$H^\pm$) is not too close to
the phase space limit, then 
there are two regimes, namely a low and a high $\tan\beta$ regime,
where the partial width of the
unconventional top quark decay becomes sizeable 
as compared to the standard decay $t\rightarrow W^+\,b$.
Nevertheless we shall focus only on the high $\tan\beta$
regime as it is this case that is correlated with the Higgs maximum rate into
the $\tau$-mode versus the hadronic mode (Cf. eq.(\ref{ratiotaucs})).
Clearly, the identification of the charged Higgs decay of the top quark
could be a matter of observing a departure from the universality prediction
for all the lepton channels through the measurement of an excess 
of inclusive (hadronic) $\tau$-events. 
However, from the non-observation of any $\tau$-lepton surplus,
one may determine
an exclusion region in the $(\tan\beta, M_{H})$-plane
\,\cite{Conway}-\cite{DPRoy} for any
(Type II) $2HDM$\,\cite{Hunter}
-- $\tan\beta$ being the  usual ratio of the two VEV's  
after SSB.
The region highlighted in this plane
consists of a sharply edged area forbidding too high values
of $\tan\beta$ in correlation with $M_H$.
In the relevant SUSY region $M_H>100\,GeV$ (see below) 
the most recent analysis would imply that values in the range
$\tan\beta\stackM 40$ would be excluded\,\cite{CDF}.

Now, an important point that we wish to make
hereafter is that the above branching ratios for this analysis could be
significantly modified by the SUSY quantum corrections. In fact,
whereas the supersymmetric effects are known to be generally small for the
standard top quark decay $t\rightarrow W^+\,b$ (Cf. Ref.\,\cite{GJSH}),
this is not necessarily so for the unconventional
mode $t\rightarrow H^+\,b$\,\cite{GJS,CGGJS}\footnote{See Ref.\cite{JSP}
for a review and a comparative study.}.  

In spite of its foreseeable importance, the impact of the SUSY 
quantum corrections on the dynamics of $t\rightarrow H^+\,b$ 
was not included in any of the aforementioned
analysis\,\cite{Conway}-\cite{DPRoy}. 
And this is especially significant
in a decay like $t\rightarrow H^+\,b$ whose sole existence could, in a sense, 
already be an indirect sign of SUSY. For, as is well-known, the CLEO
data\,\cite{CLEO} on the radiative decays $\bar{B}^0\rightarrow X_s\,\gamma$ 
(viz. $b\rightarrow s\,\gamma$) set a
rather stringent lower bound on the mass of any
charged Higgs scalar belonging to a generic $2HDM$, to wit: $M_H>240\,GeV$. 
Therefore, with only the $W^\pm$ and $H^\pm$ electroweak corrections, 
the charged Higgs mass is forced to lie in a range
where the decay $t\rightarrow H^+\,b$ becomes kinematically blocked up. 
Of course, this is so because the virtual Higgs effects go in the same
direction  
as the SM contribution. Fortunately, this situation can be remedied in the
MSSM 
where the complete formula for the $b\rightarrow s\,\gamma$
branching ratio reads (see the extensive literature\,\cite{Ng} for details):
\beq
BR (b\rightarrow s\,\gamma)\simeq BR (b\rightarrow c\,e\,\bar{\nu})\,
{(6\,\alpha_{\rm em}/\pi)\,\left(\eta^{16/23}\,A_{\gamma}+C\right)^2\over
I(m_c/m_b)\,\left[1-\frac{2}{3\pi}\,\alpha_s(m_b)f_{\rm QCD}(m_c/m_b)\right]}
\label{eq:bsg}\,,
\eeq
with
\beq
A_{\gamma}=A_{\rm SM}+A_{H^-}+A_{\chi^-\tilde{q}}
\label{eq:AS}
\eeq 
being the sum of the
SM, charged Higgs and chargino-squark
amplitudes, respectively. (The contributions from the neutralino
and gluino amplitudes are in this case generally smaller as they 
enter through FCNC.) Now,
the important feature here is that the unwanted charged Higgs effects
could to a large extent be compensated for by the chargino-stop contributions.
And in this case a relatively light charged Higgs particle would perfectly be 
allowed in the MSSM for the decay $t\rightarrow H^+\,b$ to occur.

In our renormalization framework, we use $H^{+}\rightarrow\tau^{+}\nu_{\tau}$ 
to define the parameter $\tan\beta$ through
\beq
\Gamma(H^{+}\rightarrow\tau^{+}\nu_{\tau})=
{\alpha m_{\tau}^2\,M_H\over 8 M_W^2 s_W^2}\,\tan^2\beta= 
{G_F m_{\tau}^2\,M_{H}\over 4\pi\sqrt{2}}\,\tan^2\beta\, 
(1-\Delta r^{MSSM})\,.
\label{eq:tbetainput}
\eeq
This generates a counterterm\,\cite{CGGJS}
\beq
{\delta\tan\beta\over \tan\beta}
=\frac{1}{2}\left(
\frac{\delta M_W^2}{M_W^2}-\frac{\delta g^2}{g^2}\right)
-\frac{1}{2}\delta Z_{H^\pm}
+\cot\beta\, \delta Z_{HW}+ 
\Delta_{\tau}\,,
\label{eq:deltabeta}
\eeq    
which allows to renormalize the $t\,b\,H^+$-vertex in perhaps the most
convenient way to deal with our physical process
$t\rightarrow H^+\,b\rightarrow \tau^+\,\nu_{\tau}\,b$. 
Apart from the full set of electroweak and strong corrections
from the roster of SUSY particles (squarks, 
gluinos, chargino-neutralinos and higgses), we of course include the
standard QCD correction with the running b-quark mass
evaluated at the top quark mass scale\,\cite{CD}. 
$\Delta_{\tau}$  above involves the complete MSSM effects
on $H^+\rightarrow\tau^+\,\nu_{\tau}$.

The results are presented in Figs.1-4.
We point out that in the present work we have locked together
the MSSM parameter space regions for
the two decays $b\rightarrow s\,\gamma$ and $t\rightarrow H^+\,b$
in order to find compatible solutions. In doing so
we have used the full structure involved
in eqs.(\ref{eq:bsg}),(\ref{eq:AS}). 
Notice that recently the NLO QCD effects in the SM amplitude
have been computed and the total error has diminished
from roughly $30\%$ to $15\%$ (including the error in $m_b/m_c$)
\,\cite{Misiak}.

In Fig.1 we determine the permitted region in the
$(\tan\beta,A_t)$-plane in accordance with
the CLEO data on radiative $\bar{B}^0$ decays at $2\,\sigma$. For fixed
$\mu<0$, we find that $A_t\,\mu<0$ in the allowed region. This piece of
information could be important since, as it is patent in that figure, the
trilinear coupling $A_t$ -- entering the
SUSY electroweak corrections -- becomes strongly correlated with $\tan\beta$. 
This correlation depends slightly upon the value of the charged Higgs
mass, $M_H$, and it is built-in
for the rest of the plots (Figs.2-4).
From the SM result mentioned above, we have made allowance for
an uncertainty of order $30\%$ stemming from the
non-computed NLO corrections within the MSSM.

For definiteness, and to ease comparison with the non-supersymmetric
results, we will normalize our analysis with respect
to Ref.\cite{DPRoy}. Here the $(l,\tau)$-channel, with $l$ a light
lepton, is used to search for
an excess of $\tau$-events. This should suffice to illustrate the
potential impact of the MSSM effects on this kind of physics.
To be precise, we are interested in the
$t\,\bar{t}$ cross-section for the $(l,\tau)$-channel, $\sigma_{l\tau}$,
i.e. for the final states caused by the decay sequences  
$t\,\bar{t}\rightarrow H^+\,b,W^-\,\bar{b}$ and 
$H^+\rightarrow \tau^+\,\nu_{\tau}$, $W^-\rightarrow l\,\bar{\nu}_l$ 
(and vice versa). The relevant quantity can be
easily derived from the measured value
of the canonical cross-section $\sigma_{t\bar{t}}$ for the standard
channel $t\rightarrow b\,l\,\nu_l$, $\bar{t}\rightarrow b\,q\,q'$, after
inserting appropriate branching fractions, namely 
\beq
\sigma_{l\tau}=\left[\frac4{81}\,\epsilon_1+\frac49\,
{\Gamma (t\rightarrow H\,b)\over
\Gamma (t\rightarrow W\,b)}\,\epsilon_2\right]\,\sigma_{t\bar{t}}\,,
\label{eq:bfrac}
\eeq 
where the first term in the bracket comes from the SM decay, and 
for the second term we assume (at high $\tan\beta$) $100\%$ branching 
fraction of $H^+$ into $\tau$-lepton, as explained before. Finally,
$\epsilon_i$ are detector efficiency factors\,\cite{DPRoy}. Notice that the
use of the measured value of $\sigma_{t\bar{t}}$\,\cite {Tipton}, instead 
of the predicted value within the standard NLO QCD approach\,\cite{Conto},
allows a model-independent treatment
of the result. In this respect, we note that there could be MSSM effects
on the standard mechanisms for $t\,\bar{t}$ production\,\cite{Wackeroth}
(viz. Drell-Yan $q\,\bar{q}$ annihilation and gluon-gluon fusion) as well
as corrections in the subsequent top quark decays\,\cite{GJSH}. 
 
Therefore, proceeding in this way the bulk 
of the MSSM pay-off stems from the $t\rightarrow H^+\,b$ contribution 
in eq.(\ref{eq:bfrac}). Specifically,
in Fig. 2 we determine, as a function of $\tan\beta$
and for a given Higgs mass and fixed set of SUSY parameters,
the cross-section for the $(l,\tau)$ final state.
There we show the tree-level ($\sigma_0$), QCD-corrected ($\sigma_{QCD}$)
and fully MSSM-corrected ($\sigma_{MSSM}$)
results. Of course, $\sigma_{MSSM}$ includes both the SUSY-QCD and standard
QCD effects, plus the MSSM electroweak corrections.
Note that the QCD curve is similar to the one
in Ref.\cite{DPRoy}\footnote{There
is, however, a small difference due to the fact that we use the top quark
scale, instead of the Higgs mass scale, to compute the QCD corrections.}, but
as it is also patent the full MSSM curve is quite different from the QCD
one: in fact, the two
curves lie mostly on opposite sides with respect to the tree-level curve!.

The horizontal line in Fig.2 gives the cross-section
for the number of events expected in the $(l,\tau)$-configuration
at the $95\%$ C.L. after correcting for the
detector efficiencies.
Hence the crossover points of the three curves with this
line determine (at $95\%$ C.L.) the maximum allowed value of
 $\tan\beta$ for the given set
of parameters. It is plain that the MSSM curve crosses that line
much earlier than the QCD curve, so that the
$\tan\beta$ bound is significantly tighter than in the
non-supersymmetric case. Notice that for this particular set of 
parameters the MSSM and tree-level curves turn out to meet the horizontal line
at about the same point, which means that the SUSY effects fully 
counterbalance the standard QCD correction. We remark that this
feature may occur for negative values of the higgsino mixing
parameter (in Fig.2, $\mu=-90\,GeV$). The situation with $\mu>0$ 
is different and it will be discussed below. 

In Fig.3 we present our results in the $(\tan\beta, M_H)$-plane, by
iterating the procedure followed in Fig.2 for $\mu<0$ and for charged Higgs
masses comprised in the relevant kinematical range $100\,GeV<M_H<m_t$. Here
the lower bound follows from the LEP constraint
$M_{A^0}>60\,GeV$
and the SUSY Higgs mass relations\footnote{In the absence of these
theoretical relations, we recall that the LEP absolute bound on direct
charged Higgs boson searches is much lower,
viz. $M_H>44\,GeV$\,\cite{LEPMH}.}. 
We also show the three exclusion
curves for the tree-level, QCD and MSSM corrected 
cross-sections. The excluded region in each case is the one below the curves.
By simple inspection of Fig.3,
it can hardly be overemphasized that the MSSM quantum effects
can be dramatic. Thus e.g. while for $M_H=100\,GeV$ the maximum allowed
value of $\tan\beta$ is about $46$  according to the QCD contour,
it is only about $30$ according to the MSSM contour. 
We have also checked that, after all, the modulation of the latter by the 
$b\rightarrow s\,\gamma$ constraint is not too significant even
when including the $30\%$ uncertainty mentioned above.
For, it turns out that although the branching ratio
for the $b\rightarrow s\,\gamma$ decay severely limits the set of possible 
values of $A_t$ for each $\tan\beta$ (Cf. Fig.1), the corresponding impact
on $t\rightarrow H^+\,b$ is really minor. This is due in part to
the fact that the supersymmetric electroweak corrections are {\it not} the
dominant component\,\cite{CGGJS} in $t\rightarrow H^+\,b$, and also
in part to the observed 
stabilization of its contribution within the region of parameter
space allowed by $b\rightarrow s\,\gamma$.

The above picture may undergo a significant qualitative change when we
move to the $\mu>0$ scenario. 
This can be appraised in Fig.4,
where we plot the excluded region in the $(\tan\beta, M_H)$-plane again
for the same cases as before. Although not shown, 
we have also determined the
portion of the $(\tan\beta,A_t)$-plane permitted by $b\rightarrow s\,\gamma$
for $\mu>0$ (implying that $A_t<0$), and checked that also in this
case the influence on our 
top quark analysis is not dramatic.   
The point with the $\mu>0$ scenario is that the MSSM curve is,
in contradistinction to the $\mu<0$ case, the less restrictive one. 
As a matter of fact it is even less restrictive than the original CDF 
curve for the inclusive $\tau$ channel! (Cf. Ref.\cite{Conway}). 
The reason is the following: for
$\mu>0$, the SUSY corrections have the same (negative) sign as the standard
QCD corrections and, therefore, the cross-section for the $\tau$-lepton
signal becomes extremely depleted. In Fig.4 we have chosen a heavier
SUSY spectrum than in the previous figures in order to keep the total
correction within the limits of perturbation theory. We see that
for squark masses of several hundred $GeV$ and a gluino mass of 
$1\,TeV$ the excluded area can be enforced to withdraw to a corner of 
parameter space. However, in this corner one cannot be precise any
more since further reduction would make also the Higgs sector
nonperturbative (see below).
Hence the conclusion emerging for the case $\mu>0$ is quite remarkable, to
wit: 
relatively light ($\stackM 100-120\,GeV$) charged Higgs masses
within the kinematical range of $t\rightarrow H^+\,b$ could be allowed
for essentially any admissible value of $\tan\beta$ within 
perturbation theory (i.e. $\tan\beta<60-70$).
In other words, within this scenario one could not disprove the existence
of relatively light supersymmetric charged higgses by the current methods
of $\tau$-lepton analysis at the Tevatron\,\footnote{Thus the criticism 
put forward in Ref.\cite{DPRoy}
against the large $\tan\beta$ approach\,\cite{GJS2} to
the -- nowadays residual -- $R_b$ 
anomaly would be unfounded in this case. }.

It is also interesting to compare our results with the bounds obtained 
from semileptonic and semitauonic $B$-meson decays. 
In Ref.\cite{QCDHaber} the excluded
region in the $(\tan\beta,M_H)$-plane is computed for a general $2HDM$
whereas in Ref.\cite{CJS} the corresponding MSSM analysis is performed
and it is also confronted with the (uncorrected) top quark decay exclusion
region.  
However, in the presence of the corrected results, we may compare Fig.2 of
the present work with Fig.3 of Ref.\cite{CJS} (both for $\mu<0$).
We realize that the supersymmetric results on the top quark decay greatly
improve the bound from semitauonic $B$ decays across 
the crucial region defined by $30\stackm\tan\beta\stackm 65$ and
Higgs masses ranging between $100-150\,GeV$. Even though for
$\tan\beta>65$ the semitauonic $B$-meson decays are more restrictive,
it should be pointed out that this range is already ruled out on sound
theoretical grounds, namely by the breakdown of perturbation theory; for 
instance, the top quark Yukawa coupling with the CP-odd Higgs boson $A^0$
would become $g\,m_b\,\tan\beta/2\,M_W>1$. On the other hand, the 
$\mu>0$ region is not so favoured by $B$-meson decays, but it is still
compatible with experimental data at the $1\sigma$ level for
$\tan\beta\stackm 40$\,\cite{CJS}.

In summary, from the the study of the quantum effects on the top quark
decay channel into charged Higgs particles  we arrive
at the conclusion that the
recently presented $\tau$-lepton analyses by the CDF Collaboration 
at Fermilab are in general model-dependent and could be significantly
altered by potentially underlying new physics.
In particular, since  in the absence of new interactions the results
from radiative $B$-meson decays generally preclude
the existence of charged Higgs bosons below the top quark mass, it is 
reasonable to link the existence of the decay $t\rightarrow H^+\,b$ to the 
viability of the leading candidate for physics beyond the SM, viz. the MSSM.
In this framework we find that, depending on the sign of the higgsino
mixing parameter, $\mu$, the recent $\tau$-lepton exclusion plots in 
($\tan\beta, M_H)$-space presented by CDF could
either be further strengthened or on the contrary be greatly weakened.
This dual situation could
only be decided from additional experimental information unambiguously 
favouring a given sign of $\mu$ in other physical processes. 
We remark that although for brevity sake we have presented our numerical
analysis for a given choice of the MSSM parameters, we have checked that
our conclusions hold basically unaltered in ample regions of parameter space
involving typical sparticle masses of a few hundred
$GeV$\,\cite{GuaschThesis}. 
While the details of the exclusion plot in $(\tan\beta,M_H)$-space may
depend on the particular channel used to tag a potential
excess of $\tau$-leptons, all of these plots (and of course also 
the one from the inclusive measurement) should undergo significant changes.
Finally, it is clear that similar considerations 
apply to experiments of the same nature being planned for
the future at the LHC. Thereby a general conclusion 
seems to consolidate\,\cite{JSP}:
In contrast to gauge boson observables, the MSSM
quantum effects on Higgs boson physics can be rather large and
should not be ignored in future searches at the Tevatron and 
at the LHC.


{\bf Acknowledgements}:
\noindent
J.S. thanks W. Hollik for discussions and for the warm hospitality at 
the Institut f{\"u}r Theoretische Physik der Univ. Karlsruhe where this work
was finished. He also thanks W. de Boer for useful remarks. 
Financial support of J.S. by the Spanish Ministerio de Eduaci{\'o}n y
Ciencia through the Programa de Acciones Integradas (Acci{\'o}n No. HA1996-0030) 
is gratefully acknowledged. J.G. has been financed by a 
grant of the Comissionat per a Universitats i Recerca, Generalitat de 
Catalunya. This work has also been partially supported by CICYT 
under project No. AEN95-0882.


\vspace{1cm}
\begin{center}
\begin{Large}
{\bf Figure Captions}
\end{Large}
\end{center}
\begin{itemize}
\item{\bf Fig.1}  The allowed region (shaded area) in the
 $(A_t,\tan\beta)$-plane by the $b\rightarrow s\,\gamma$ decay within
the framework of the MSSM, and
for a given set of inputs. Here the most relevant SUSY parameters are:
the gaugino-higgsino mass parameters $M$ and $\mu$,
the lightest stop and sbottom masses $m_{\tilde{t_1},\tilde{b_1}}$, the 
trilinear couplings $A_{t,b}$, and the gluino mass $m_{\tilde{g}}$. 
The corresponding parameters for the other squark and slepton generations are 
also given, and the Higgs mass is fixed at $M_H=120\,GeV$.  

\item{\bf Fig.2} The cross-section for the $(\tau,l)$-channel (in fb)
for the tree-level ($\sigma_0$), QCD-corrected ($\sigma_{QCD}$)
and fully MSSM-corrected ($\sigma_{MSSM}$) cases, for the same parameters
as in Fig.1. The horizontal line gives the $95\%$ C.L. cross-section for the 
observation of the $(\tau,l)$ final state.

\item{\bf Fig.3} The $95\%$ C.L. exclusion plot in the $(\tan\beta,
  M_H)$-plane 
for $\mu<0$. Shown are the tree-level (dashed), QCD-corrected
(dotted) and fully 
MSSM-corrected (continuous) contour lines.
The excluded region in each case is the one lying below these curves.
The set of parameters is as in Fig.1, with $A_t$ within the allowed region. 

\item{\bf Fig.4} As in Fig.3, but for a $\mu>0$ scenario characterized by a 
heavier SUSY spectrum.

\end{itemize}

\newpage

\thispagestyle{empty}

\centering{\mbox{\epsfig{width=15cm,height=10cm,file=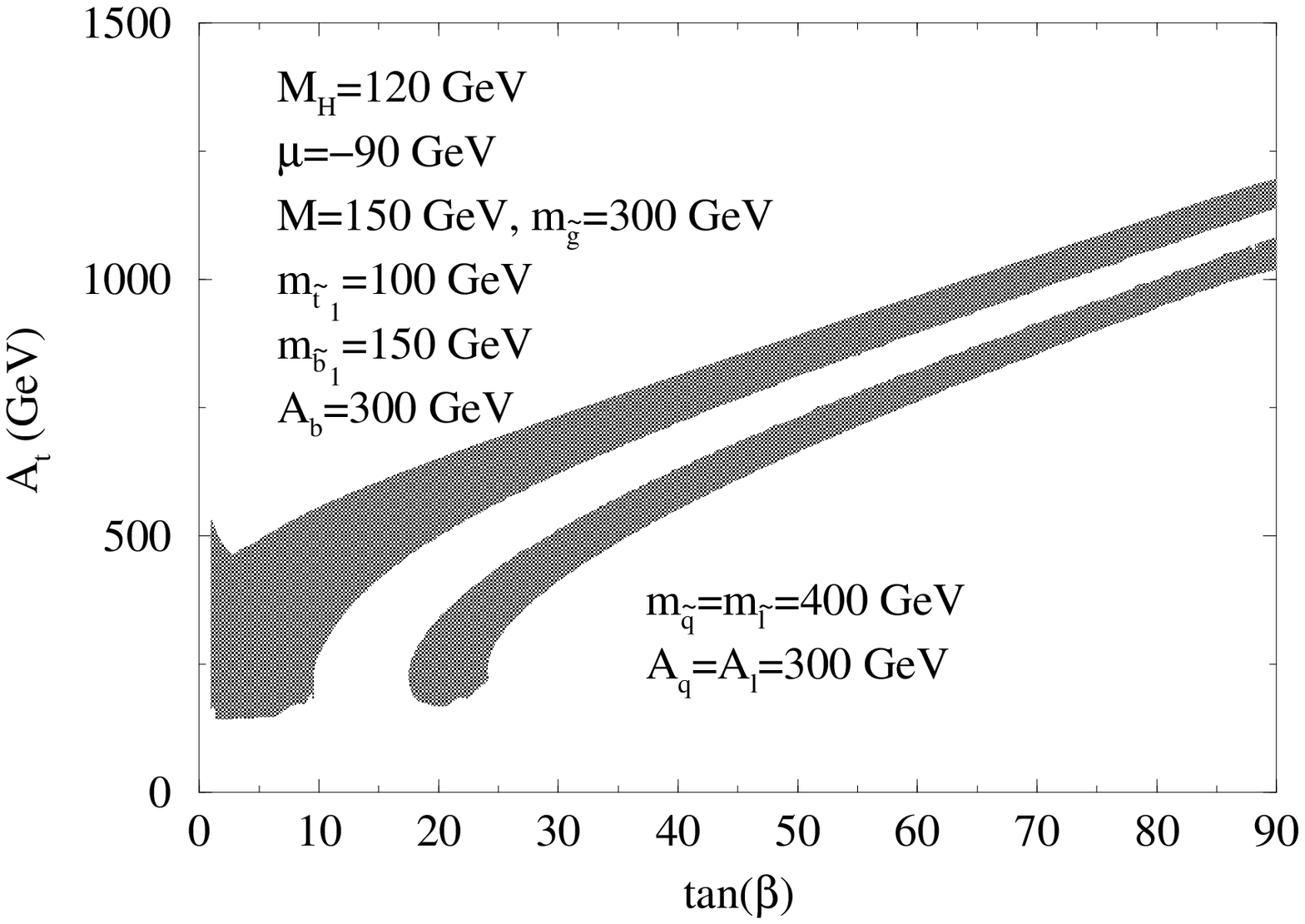}}\\
{\Large Fig.1}}
\thispagestyle{empty}

\centering{\mbox{\epsfig{width=15cm,height=10cm,file=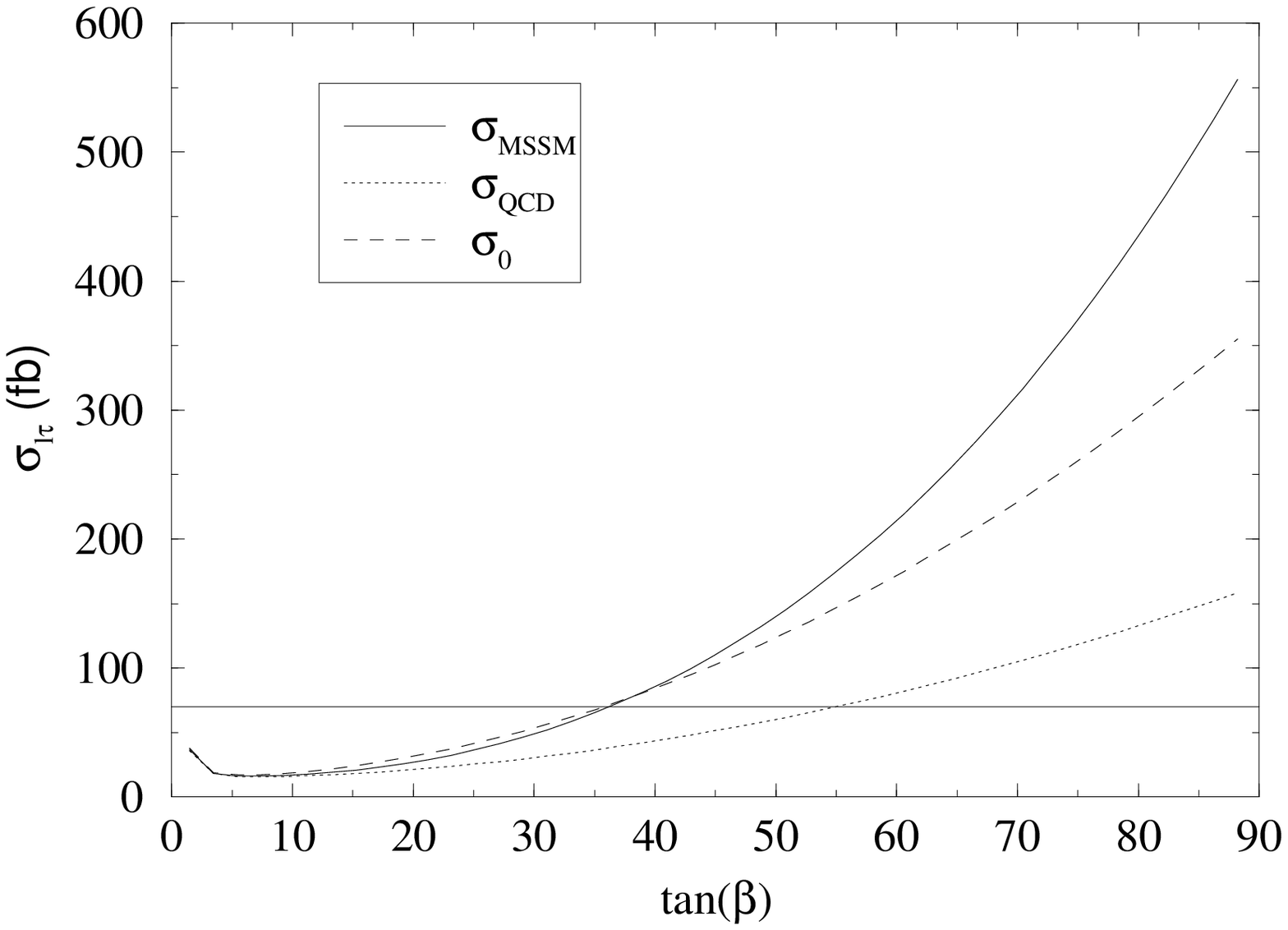}}\\
{\Large Fig.2}}
\thispagestyle{empty}

\centering{\mbox{\epsfig{width=15cm,height=10cm,file=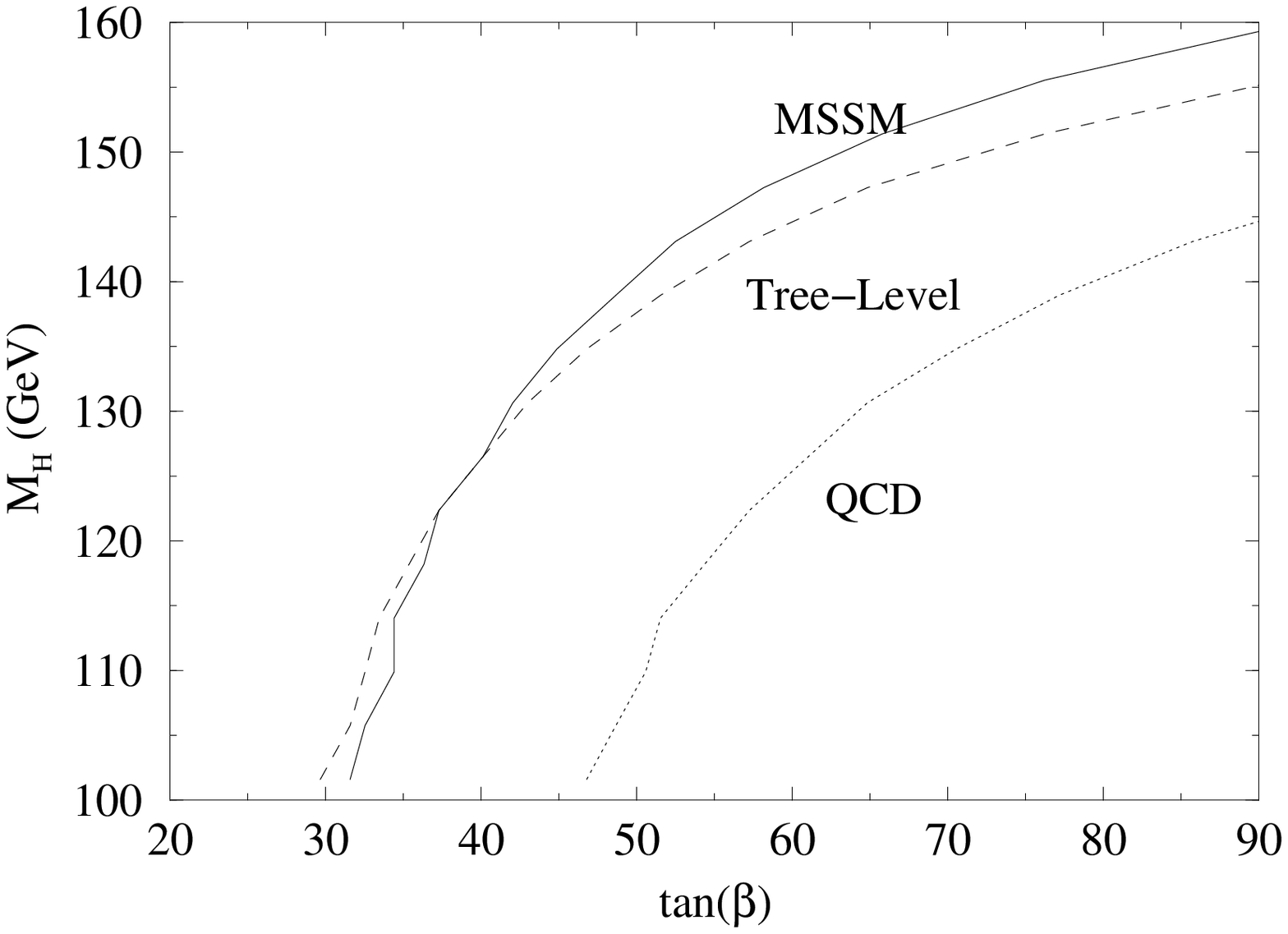}}\\
{\Large Fig.3}}
\thispagestyle{empty}

\centering{\mbox{\epsfig{width=15cm,height=10cm,file=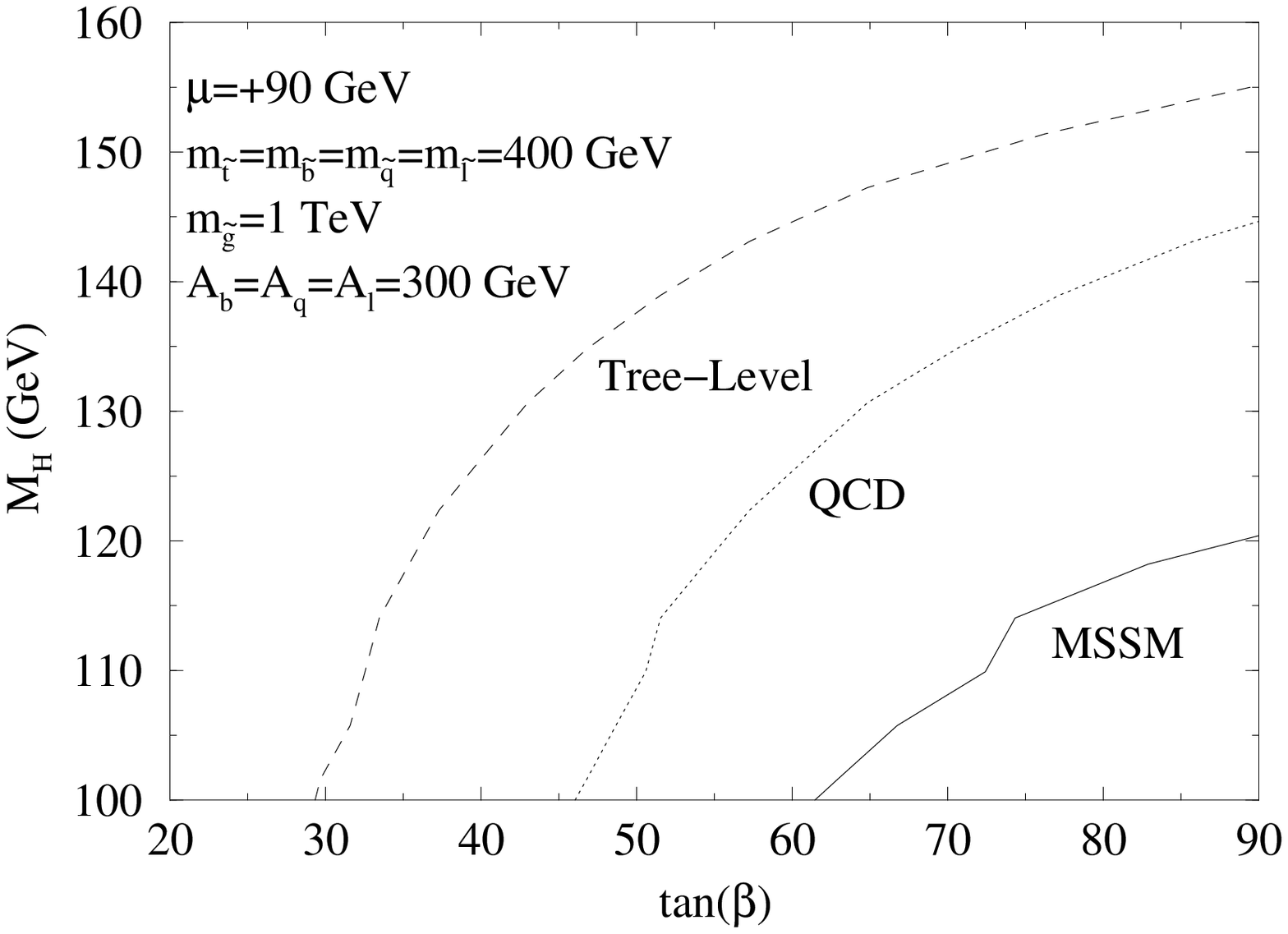}}\\
{\Large Fig.4}}

\end{document}